\pdfoutput=1

\documentclass[
aps,prl,
reprint,
a4paper,
superscriptaddress,
longbibliography,
preprintnumbers,
]{revtex4-1}
\usepackage[utf8]{inputenc}
\usepackage[T1]{fontenc}

\usepackage{amsmath,amsthm,amsfonts}
\usepackage{braket}
\usepackage{graphicx}
\usepackage{hyperref}
\usepackage{gensymb}
\usepackage{bbm}
\usepackage{color}
\usepackage{booktabs}
\usepackage{siunitx}

\hypersetup{colorlinks=true}
\hypersetup{citecolor=blue}
\hypersetup{urlcolor=blue}

\begin{document}

\title{Experimental certification of an informationally complete quantum measurement in a device-independent protocol}


\author{Massimiliano Smania}
\email{massimiliano.smania@fysik.su.se}
\affiliation{Department of Physics, Stockholm University, S-10691 Stockholm, Sweden}

\author{Piotr Mironowicz}
\email{piotr.mironowicz@gmail.com}
\affiliation{Department of Algorithms and System Modeling, Faculty of Electronics, Telecommunications and Informatics, Gda\'nsk University of Technology}
\affiliation{National Quantum Information Centre in Gda\'nsk, Sopot 81-824, Poland}
\affiliation{International Centre for Theory of Quantum Technologies, University of Gda\'nsk, Wita Stwosza 57, 80-308 Gda\'nsk, Poland}

\author{Mohamed Nawareg}
\affiliation{Department of Physics, Stockholm University, S-10691 Stockholm, Sweden}

\author{Marcin Paw\l{}owski}
\affiliation{National Quantum Information Centre in Gda\'nsk, Sopot 81-824, Poland}
\affiliation{International Centre for Theory of Quantum Technologies, University of Gda\'nsk, Wita Stwosza 57, 80-308 Gda\'nsk, Poland}

\author{Ad\'an Cabello}
\affiliation{Departamento de F\'{\i}sica Aplicada II, Universidad de Sevilla, E-41012 Sevilla, Spain}
\affiliation{Instituto Carlos~I de F\'{\i}sica Te\'orica y Computacional, Universidad de Sevilla, E-41012 Sevilla, Spain}

\author{Mohamed~Bourennane}
\email{boure@fysik.su.se}
\affiliation{Department of Physics, Stockholm University, S-10691 Stockholm, Sweden}

\begin{abstract}
    Minimal informationally complete positive operator-valued measures (MIC-POVMs) are special kinds of measurement in quantum theory in which the statistics of their $d^2$-outcomes are enough to reconstruct any $d$-dimensional quantum state. For this reason, MIC-POVMs are referred to as standard measurements for quantum information. Here, we report an experiment with entangled photon pairs that certifies, for what we believe is the first time, a MIC-POVM for qubits following a device-independent protocol (i.e., modeling the state preparation and the measurement devices as black boxes, and using only the statistics of the inputs and outputs). Our certification is achieved under the assumption of freedom of choice, no communication, and fair sampling.
\end{abstract}

\maketitle


\section{Introduction}
A minimal informationally complete positive operator-valued measure (MIC-POVM) \cite{RBSC04,Weigert06} is a measurement on a $d$-dimensional quantum system that: (i) is informationally complete (IC), i.e., its statistics determine completely any quantum state and allow for a simple state-reconstruction and (ii) is minimal, since it has the minimum number of outcomes a measurement must have to be IC, namely, $d^2$ \cite{CDS07}. MIC-POVMs are fundamental in quantum information. For example, they are crucial for optimal quantum state tomography \cite{CFS02,REK04}, quantum key distribution with optimal trade-off between security and key rate \cite{FS03}, device-independent certification of optimal randomness from one bit of entanglement \cite{APVW16,OBDC18}, and optimal entanglement detection \cite{SAZG18}. Arguably, MIC-POVMs are the ``standard'' measurements in quantum information \cite{Fuchs02} and thus have a privileged role in information-theoretic reconstructions of quantum theory \cite{Fuchs02}. 

Experimentally, MIC-POVMs have been aimed at in photonic experiments of qubit \cite{DKLL08}, qutrit \cite{MTSTFS11,PMMVDSP13}, and two-qubit \cite{HTSZLYWXLG18} tomography, quantum key distribution \cite{DKLL08}, and generalized measurements using quantum walks \cite{HTSZLYWXLG18,ZYKXLG15}. However, all these experiments made assumptions about the state preparation and the functioning of the measurement devices and therefore have limited applicability to cryptographic scenarios. 

Here, we address the problem of experimentally certifying a MIC-POVM for qubits following a ``device independent'' (DI) protocol \cite{APVW16,OBDC18,ABGMPS07,AM16}. That is, modeling the state preparation and the measurement devices as black boxes and using only the statistics of the inputs and outputs obtained within a Bell inequality experiment. The idea behind the experiment is to integrate the MIC-POVM within a Bell inequality experiment and use it to produce correlations that, according to quantum theory, are only attainable with a MIC-POVM for qubits. 
Our results not only allow us to certify a MIC-POVM for qubits in a DI protocol, but also constitute the first experimental observation of ``qubit correlations which can only be explained by four-outcome non-projective measurements'' \cite{GGGCBDXCKVL16}.

\section{Certification Methods}

\begin{figure}[htbp]
\centering\includegraphics[width=.95\linewidth]{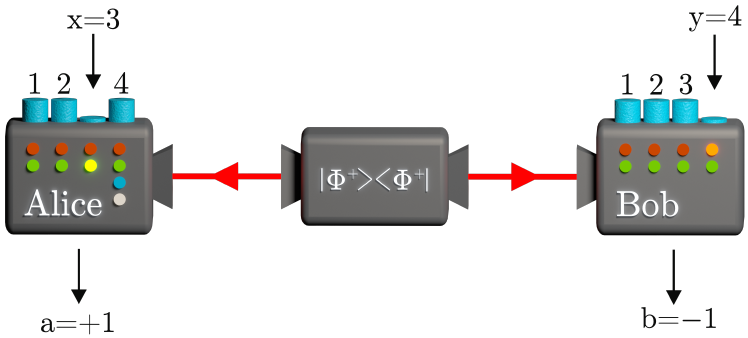}
\caption{The scenario considered in our experiment consists of two parties, Alice and Bob, who perform local measurements on maximally entangled pairs of qubits. See further details in the text.}
\label{fig:scheme}
\end{figure}

To certify a four-outcome MIC-POVM in a DI way, we use the bipartite Bell scenario shown in Fig.~\ref{fig:scheme}. There, in the middle, is a source emitting pairs of particles. One of the particles is measured by one party, Alice, and the other particle is measured by the other party, Bob. Alice chooses her measurement from a set of four measurements: three two-outcome measurements $A_x$, with $x \in \{1,2,3\}$, and one four-outcome measurement $A_4$. Bob chooses his measurement from a set of four two-outcome measurements $B_y$, with $y \in \{1,2,3,4\}$. The possible outcomes of the two-outcome measurements are $+1$ and $-1$, and the possible outcomes of the four-outcome measurements are $1,2,3$, and $4$. The outcomes of $A_x$ and $B_y$ are denoted by $a$ and $b$, respectively. From the experimental results, we obtain the set of conditional probabilities $\{P(a,b|x,y)\}$.

In our experiment, we are interested in the conditional probabilities appearing in the following Bell inequality introduced in Ref.~\cite{APVW16}:
\begin{eqnarray} \label{eq:el_mod}
\beta^\text{m}_\text{el}=\beta_\text{el} - k \sum^4_{i=1} P(a=i,b=+1|x=4,y=i),
\end{eqnarray}
where $\beta_\text{el}$ is the Bell operator of the so-called ``elegant Bell inequality'' \cite{Gisin09,APVW16,OBDC18}, namely,
\begin{equation} \label{eq:el}
\begin{aligned}
\beta_\text{el} &= E_{11}+E_{12}-E_{13}-E_{14}+E_{21}-E_{22}\\
&+E_{23}-E_{24}+E_{31}-E_{32}-E_{33}+E_{34},
\end{aligned}
\end{equation}
where $E_{xy} = \sum_{a,b} a b P(a,b|x,y)$. For local hidden variable theories, $\beta_\text{el}$ is upper-bounded by $6$. In contrast, in quantum theory $\beta_\text{el}$ is upper-bounded by $4 \sqrt{3} \approx 6.928$. The quantum maximum can be achieved with pairs of qubits in state $|\Phi^+\rangle = \frac{1}{\sqrt{2}} \left( |00\rangle + |11\rangle \right)$ and the following projective measurements:
\begin{equation} \label{eq:proj}
\begin{aligned}
A_1&= \sigma_x, \;\;& B_1&=\frac{1}{\sqrt{3}}(\sigma_x-\sigma_y+\sigma_z),\\
A_2&= \sigma_y, \;\; & B_2&=\frac{1}{\sqrt{3}}(\sigma_x+\sigma_y-\sigma_z),\\
A_3&= \sigma_z, \;\;& B_3&=\frac{1}{\sqrt{3}}(-\sigma_x-\sigma_y-\sigma_z),\\
& & B_4 & =\frac{1}{\sqrt{3}}(-\sigma_x+\sigma_y+\sigma_z), 
\end{aligned}
\end{equation}
where $\sigma_i$ are the Pauli matrices.

According to quantum theory, our target Bell operator $\beta^\text{m}_\text{el}$ is also upper-bounded by $4\sqrt{3}$. This quantum maximum can be attained with state $|\Phi^+\rangle$ and the measurements in Eq.~\eqref{eq:proj}. However, in this case, the second term in Eq.~\eqref{eq:el_mod} is zero if and only if $A_4$ is a qubit symmetric MIC-POVM whose elements are anti-aligned with Bob's measurements $B_y$ in Eq.~\eqref{eq:proj}. That is, if $A_4$ is the four-outcome POVM, defined by the following elements:
\begin{equation} \label{eq:povm}
\begin{aligned}
A_{4,1}&=\frac{1}{2}\begin{pmatrix} \alpha&-\beta(1+i)\\ \beta(-1+i)&1-\alpha \end{pmatrix},\\
A_{4,2}&=\frac{1}{2}\begin{pmatrix} 1-\alpha&\beta(-1+i)\\ -\beta(1+i)&\alpha \end{pmatrix},\\
A_{4,3}&=\frac{1}{2}\begin{pmatrix} 1-\alpha&\beta(1-i)\\ \beta(1+i)&\alpha \end{pmatrix},\\
A_{4,4}&= \frac{1}{2}\begin{pmatrix} \alpha&\beta(1+i)\\ \beta(1-i)&1-\alpha \end{pmatrix},
\end{aligned}
\end{equation}
where $\alpha= \frac{3-\sqrt{3}}{6}$ and $\beta= \frac{\sqrt{3}}{6}$. In this case, the extremes of the four unit vectors associated to the elements of $A_4$ define a regular tetrahedron within the Bloch sphere. 

Any measurement different than the one defined in Eq.~\eqref{eq:povm} will produce a smaller value for $\beta^\text{m}_\text{el}$. While certifying a \textit{symmetric} MIC-POVM requires ideal conditions, we can use the property above to test whether a genuine four-outcome, MIC-POVM has indeed been realized in the experiment, by computing the maximum of $\beta^\text{m}_\text{el}$ that can be obtained using three-outcome measurements. In order to do this, let us consider the following generalization of $\beta^\text{m}_\text{el}$:
\begin{equation}
\label{eq:modified_el}
\begin{split}
\sum_{x=1}^3 \sum_{y=1}^4 \gamma_{xy} E_{xy} - k \sum_{y=1}^4 \sum_{a=1}^4 \sum_{b=\pm 1} \gamma_{bxy} P(a,b|4,y).
\end{split}
\end{equation}
We compute the maximum value of Eq.~\eqref{eq:modified_el} that can be obtained using three-outcome measurements. That is, the maximum value of
\begin{multline}\label{eq:modified_el_3outcomes}
\max_{j=1,2,3,4} \left[ \sum_{x=1}^3 \sum_{y=1}^4 \gamma_{xy} E_{xy} \right. \\
\left. - k \sum_{y=1}^4 \sum_{a \neq j} \sum_{b=\pm 1} \gamma_{bxy} P(a,b|4,y) \right].
\end{multline}
Each of the maximizations within Eq.~\eqref{eq:modified_el_3outcomes} are taken with a constraint that the $j$th outcome of Alice's $A_4$ measurement has probability $0$.
The larger the gap between the experimental value of Eq.~\eqref{eq:modified_el} and the maximum possible value of Eq.~\eqref{eq:modified_el_3outcomes}, the more confident we can be that indeed a genuine four-outcome POVM has been performed.

Most crucially, as we show in the Supplemental Material, four irreducible outcomes in dimension $2$ imply information completeness, which is arguably the most important feature of a quantum measurement.

Finally, while Bell scenarios can in general be useful for measurement certification, it is important to point out that the Bell inequality in the protocol above is tailored to the specific measurement targeted in our certification. In order to use the same procedure for an arbitrary measurement, a different Bell inequality would in general be required. 
In fact, finding the optimal Bell inequality for certifying in a device-independent protocol a given generalized measurement is, in general, a difficult problem.

\section{Experiment}
\subsection{Experimental setup}

\begin{figure*}
    \includegraphics[width=.95\linewidth]{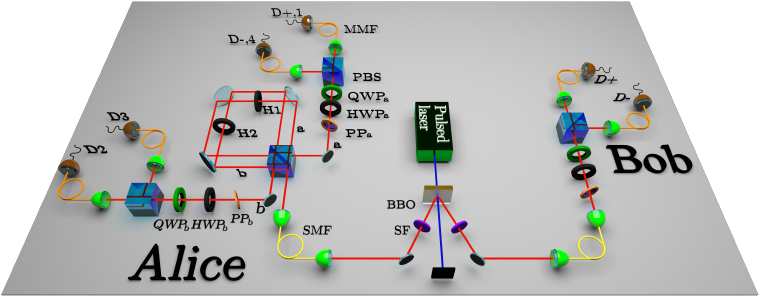}
	\caption{Experimental setup. The following components are used: a beta-barium borate non-linear crystal (BBO), \SI{3}{nm} narrow spectral filters (SF), single-mode optical fibers (SMF), phase plates (PP), lambda-half wave plates (HWP), lambda-quarter wave plates (QWP), polarizing beam splitters (PBS), multi-mode optical fibers (MMF), and single-photon detectors (DET).}
	\label{fig:setup}
\end{figure*}

A type-I spontaneous parametric down-conversion source is used to generate entangled photon pairs in state $\ket{\Phi ^+} = 1/ \sqrt{2} (\ket{HH}+\ket{VV})$, where $H$ and $V$ denote horizontal and vertical polarization, respectively. Pairs of entangled photons at \SI{780}{nm} are produced in two orthogonally oriented \SI{2}{mm} thick beta barium borate (BBO) crystals, pumped with a femto-second laser at \SI{390}{nm}. As shown in Fig.~\ref{fig:setup}, these photons go through \SI{1}{nm} spectral bandpass filters (SF), and are then coupled into single-mode fibers (SMF) to have perfect spatial mode overlap between the two polarizations. These SMFs then bring the photons to Alice's and Bob's measurement stations. Whenever projective measurements are performed on both sides (i.e., whenever $x \in \{1, 2, 3\}$ and $y \in \{1, 2, 3, 4\}$), the two measurement stations are identically composed by a lambda-half wave plate (HWP), a lambda-quarter wave plate (QWP), and a polarization beam-splitter (PBS). Multi-mode fibers (MMF) finally collect the photons at the four outcomes and bring them to the single-photon avalanche photodiodes (APDs). In addition, Bob's station includes a phase plate (PP).

In order to implement the four-outcome POVM, Alice's measurement station couples the two-dimensional polarization space with a counter-propagating two-path Sagnac interferometer, which makes transformations in an effectively four-dimensional space possible using two HWPs.
At the two outputs of the interferometer, PPs, HWPs and QWPs are used in combination with PBSs to perform the POVM (for more details, see Supplemental Material). MMFs connected to APDs again gather photons at the four outcomes. Detection counting is performed with a coincidence unit (CU) using \SI{1.6}{ns} coincidence windows.

\begin{figure}[htbp]
\centering\includegraphics[width=\linewidth]{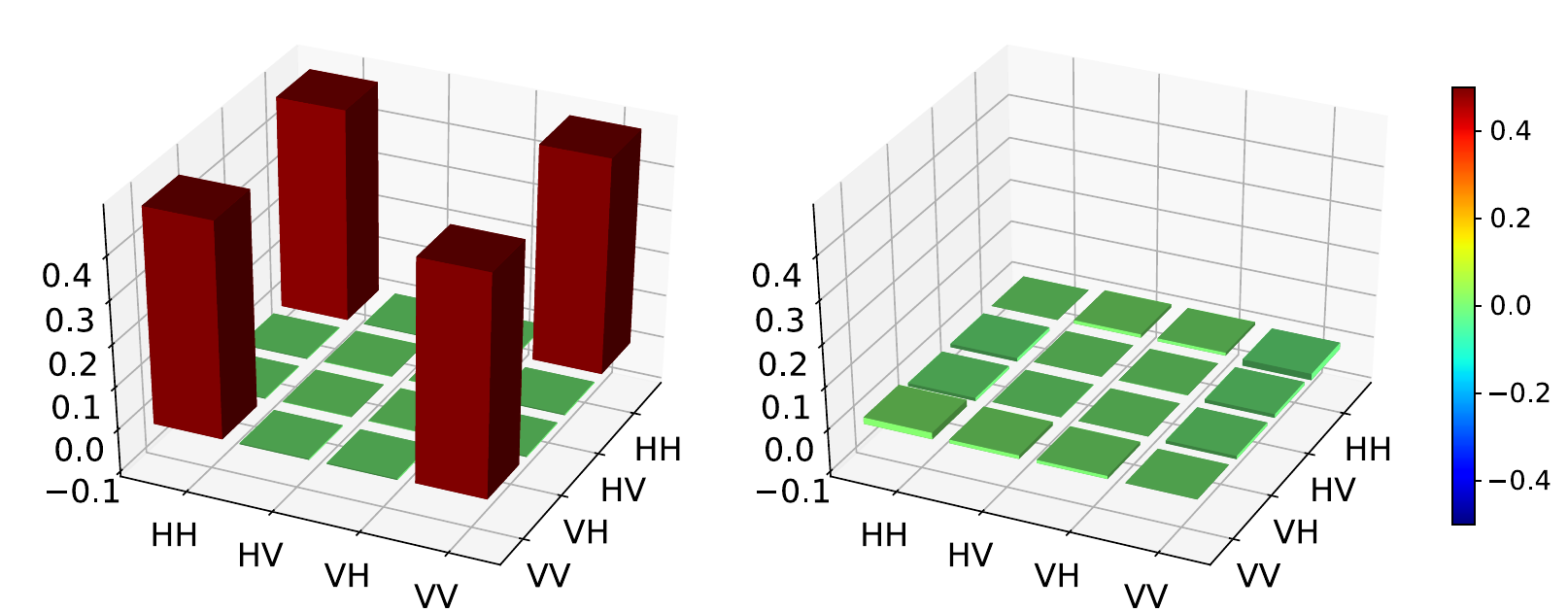}
	\caption{Tomography of the prepared maximally entangled state. Real (left) and imaginary (right) parts.}
	\label{fig:tomography}
\end{figure}

A two-photon rate of about $150$ coincidences per second was maintained throughout the experiment. Each measurement setting lasted 30 seconds, and all measurements were repeated a total of $23$ times. This was done in order to decrease the repeatability error of the motorized wave plate mounts.

\subsection{Results}
The maximization of Eq.~\eqref{eq:modified_el_3outcomes} for fixed coefficients $\gamma_{xy}$ and $\gamma_{bxy}$ should be made over the set of probabilities allowed by quantum theory. To obtain an upper bound on its value, we use the semi-definite programming method of Navascu\'es, Pironio, and Ac\'{\i}n (NPA) \cite{NPA08} implemented in the Python package Ncpol2spda \cite{Wittek15}. 

To identify the values of $\gamma_{xy}$ and $\gamma_{bxy}$ that provide the largest gap between the experimental value of~\eqref{eq:modified_el} and the maximum possible value of~\eqref{eq:modified_el_3outcomes}, we performed a series of numerical optimizations using the Nelder-Mead method \cite{NM65}, with target function defined as the difference between experimental value of~\eqref{eq:modified_el} and a bound of~\eqref{eq:modified_el_3outcomes}, with variable coefficients $\gamma_{xy}$ and $\gamma_{bxy}$ and fixed values of $k$. As a result, we obtained the following Bell operator:
\begin{widetext}
	\begin{equation}\label{eq:optimised_el}
	\begin{aligned}
	& 0.9541 E_{11} + 0.9917 E_{12} - 0.9767 E_{13} - 1.0064 E_{14} \\
	&+ 0.9514 E_{21} - 0.9921 E_{22} + 0.8211 E_{23} - 1.0237 E_{24} \\
	&+ 1.0641 E_{31} - 1.0044 E_{32} - 1.0579 E_{33} + 1.1563 E_{34} \\
	&-3[1.2068 P(1,1|4,1) -0.0374 P(1,2|4,1) -0.0034 P(2,1|4,1) +0.0140 P(2,2|4,1) \\
	&+0.0006 P(3,1|4,1) +0.0268 P(3,2|4,1) -0.0163 P(4,1|4,1) -0.0155 P(4,2|4,1) \\
	&-0.0033 P(1,1|4,2) +0.0184 P(1,2|4,2) +1.1156 P(2,1|4,2) -0.0046 P(2,2|4,2) \\
	&-0.0125 P(3,1|4,2) +0.0401 P(3,2|4,2) -0.0175 P(4,1|4,2) -0.0240 P(4,2|4,2) \\
	&-0.0108 P(1,1|4,3) +0.0153 P(1,2|4,3) -0.1195 P(2,1|4,3) +0.1752 P(2,2|4,3) \\
	&+0.6201 P(3,1|4,3) +0.0149 P(3,2|4,3) -0.0399 P(4,1|4,3) +0.0527 P(4,2|4,3) \\
	&+0.0058 P(1,1|4,4) -0.0149 P(1,2|4,4) +0.0025 P(2,1|4,4) +0.0205 P(2,2|4,4) \\
	&+0.0150 P(3,1|4,4) +0.0212 P(3,2|4,4) +0.9565 P(4,1|4,4) -0.0023 P(4,2|4,4)].
	\end{aligned}
	\end{equation}
\end{widetext}
The upper bounds on the maximum possible value of $\beta_\text{IC}$ in Eq.~\eqref{eq:optimised_el}, obtained using the third level of the NPA method, in case of three-outcome measurements and in case of any quantum measurement are:
\begin{equation}
    \beta_\text{IC} \overset{\text{3-outcome}}{\leq} 6.8782 \overset{\text{Quantum}}{\leq} 6.9883,
\end{equation}
whereas our experimental result is:
\begin{equation}
    \beta^\text{exp}_\text{IC} = 6.960 \pm 0.007
\end{equation}
(more detailed results are provided in Tables \ref{tab:proj} and \ref{tab:povm}). Therefore, the experimental value violates the three-outcome bound by more than $11$ standard deviations, certifying that Alice's $A_4$ measurement was indeed an irreducible four-outcome measurement, under the assumption that the system at Alice's laboratory is a qubit. 
In order to remove this assumption, we used the SWAP method \cite{YVBSN14} to calculate the two-qubit state fidelity with the maximally entangled Bell state $|\Phi^+\rangle$ certified by the experimental data contained in Tab.~\ref{tab:proj} for $\beta_\text{el}$ in Eq.~\eqref{eq:el}. To this end we employed the level $3+AABB$ of the NPA hierarchy~\cite{NPA08}. The resulting fidelity was $0.947$.

This means that a qubit measurement occurs in at least $94.7\%$ of times. The only alternative to a qubit MIC-POVM is that in $0.947$ of the cases a three-outcome measurement on a qubit was used and in the remaining $0.053$ of the cases a four-outcome measurement on a higher-dimensional system was used. However, in such a case, the maximal possible value to be observed is not greater than $0.947 \times 6.8782 + 0.053 \times 6.9883 \approx 6.8840$, which is smaller than the experimental value, namely, $6.960$. Similarly, one can calculate that the critical fidelity $\eta_{crit}$ to two dimensional state for MIC-POVMs is given by:
$(6.9883 - \beta_{exp})/(6.9883 - 6.8782) \approx 0.257$.

Even though our DI protocol relies on this method, we can provide additional, non DI arguments, which suggest that the actual state fidelity was considerably higher.
Firstly, we tested the quality of the polarization entanglement by performing a complete nine-measurement state tomography of the Alice-Bob system. The tomography of the joint state is shown in Fig.~\ref{fig:tomography}.
The experimental fidelity with state $|\Phi^+\rangle$ was near optimal at $(99.6 \pm 0.1)\%$. 
Secondly, our measured value of the Bell operator of the Elegant Bell Inequality was $\beta_\text{el} = 6.909 \pm 0.007$, which corresponds to $99.7\%$ of the quantum bound and is less than three standard deviations away from it. As mentioned above, a result equal to the quantum bound would imply that the joint state is a maximally entangled qubit-qubit state \cite{APVW16,OBDC18}.

As a final remark, although our protocol is DI, we have assumed freedom-of-choice, fair-sampling and no-communication in our experiment. As we show in the Supplemental Material, closing the detection loophole would require overall system efficiencies above $94\%$, outside the reach of state of the art photonics experiments.

\begin{table}[htbp]
	\begin{center}
		\begin{tabular}{ccc}
			\hline
			Setting\; & Theory & Experiment \\ \hline
			$E_{11}$ & $1/\sqrt{3}\approx0.577$ & $0.553 \pm 0.002$ \\
			$E_{12}$ & $0.577$ & $0.573 \pm 0.002$ \\ 
			$E_{13}$ & $-0.577$ & $-0.581 \pm 0.002$ \\  
			$E_{14}$ & $-0.577$ & $-0.543 \pm 0.002$ \\  
			$E_{21}$ & $0.577$ & $0.589 \pm 0.002$ \\  
			$E_{22}$ & $-0.577$ & $-0.599 \pm 0.002$ \\ 
			$E_{23}$ & $0.577$ & $0.529 \pm 0.002$ \\  
			$E_{24}$ & $-0.577$ & $-0.579 \pm 0.002$ \\ 
			$E_{31}$ & $0.577$ & $0.584 \pm 0.002$\\  
			$E_{32}$ & $-0.577$ & $-0.557 \pm 0.002$ \\  
			$E_{33}$ & $-0.577$ & $-0.621 \pm 0.002$ \\ 
			$E_{34}$ & $0.577$ & $0.601 \pm 0.002$\\ 
			$\beta_\text{el}$ & $4 \sqrt{3}\approx6.928$ & $6.909 \pm 0.007$\\ \hline
		\end{tabular}
	\end{center}
\caption{\label{tab:proj}Experimental values for the combinations of settings needed to test the elegant Bell inequality.}
\end{table}


\begin{table}[htbp]
	\begin{center}
		\begin{tabular}{ccc}
			\hline
			$P(a=i,b=+1|x=4,y=i)$\;\; & Theory & Experiment \\ \hline
			$P(1,+1|4,1)$ & $0$ & $0.0021 \pm 0.0001$ \\ 
			$P(2,+1|4,2)$ & $0$ & $0.0020 \pm 0.0001$ \\ 
			$P(3,+1|4,3)$ & $0$ & $0.0025 \pm 0.0001$ \\ 
			$P(4,+1|4,4)$ & $0$ & $0.0025 \pm 0.0001$ \\ 
			Sum & $0$ & $0.0091 \pm 0.0002$\\ \hline
		\end{tabular}
	\end{center}
\caption{\label{tab:povm}Experimental values for the probabilities of the outcomes of the MIC-POVM that are most relevant to the DI certification protocol [see Eq.~\eqref{eq:el_mod}].}
\end{table}

\subsection{State tomography with the MIC-POVM}

\begin{figure}[htbp]
\centering\includegraphics[width=\linewidth]{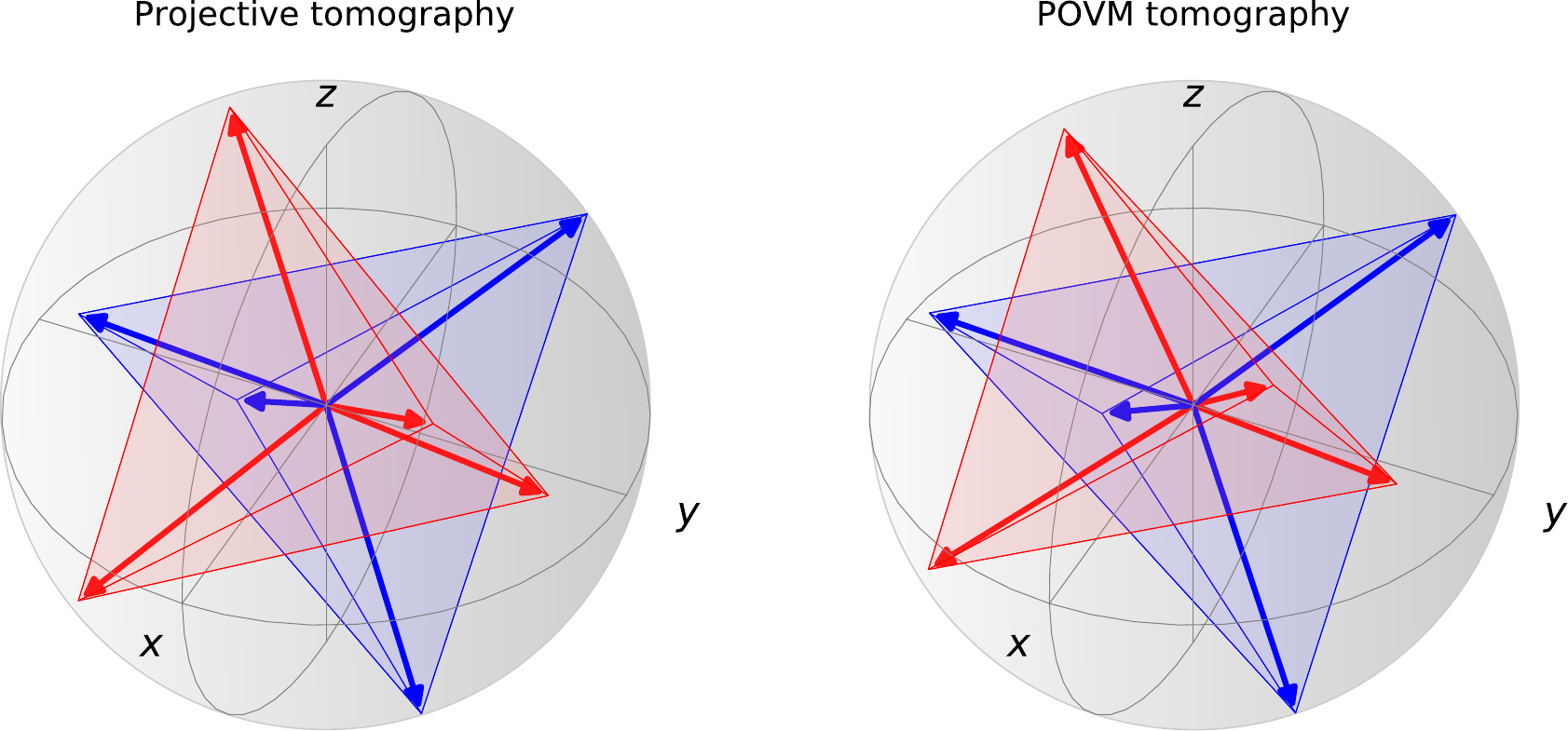}
	\caption{Reconstruction of eight Alice's local qubit states, conditioned on Bob's setting and outcome, as obtained from standard projective tomography (left) and MIC-POVM tomography (right).}\label{fig:comparison_tomo}
\end{figure}

In order to test the tomographic capabilities of our certified MIC-POVM against the standard tomographic methods based on projective measurements, we reconstructed eight different Alice's local qubit states (those naturally occurring in our Bell scenario when we condition Alice's state to Bob's measurements and results) using both methods. Firstly, a standard tomographic analysis from the experimental statistics of {\em three} projective measurements (in our case, $A_1,A_2$, and $A_3$; that is, $\sigma_x,\sigma_y$, and $\sigma_z$). Secondly, using only the experimental statistics of our {\em single} four-outcome measurement. The resulting reconstructed local states should be identical for both methods and, ideally, must point to the corners of a regular tetrahedron in the Bloch sphere. 
In the case of the MIC-POVM, a simple formula connects the four experimental frequencies produced by the single measurement with the tomographic reconstruction:
\begin{equation}
 \Vec{s}=3 \sum_{j=1}^4 f_j \Vec{A}_j,
\end{equation}
where $\Vec{s}$ is the unknown Bloch vector, each $\Vec{A}_j$ is one of the four elements of the symmetric MIC-POVM set (see Eq.~\eqref{eq:povm}), and $f_j$ is its corresponding measured frequency \cite{REK04}.
On the other hand, the six experimental frequencies provided by the three projective measurements were used, through linear inversion \cite{Schwemmer15}, to reconstruct the same states.
The results of both methods are presented in Fig.~\ref{fig:comparison_tomo}.

The fidelity between corresponding vectors was in all eight cases equal to or greater than $99.5\%$, indicating that the two tomographic methods yielded near-optimally overlapping results, and the four-outcome POVM is informationally complete and effectively symmetric. More detailed results are provided in the Supplemental Material.

All the experimental uncertainties reported were calculated with \textit{a priori} evaluation of known sources of error, and subsequent propagation in the results. The sources of errors included in our analysis were: photon counting statistics, precision of wave plate rotation, detector dark counts. and higher order down-conversion events (see Supplemental Material for more details).

\section{Conclusions}
Quantum information identifies MIC-POVMs as the standard quantum measurements for information processing tasks as they are informationally complete and optimal for tomographic and cryptographic purposes. On the other hand, the device-independent paradigm provides the arguably optimal way to test quantum devices, as it reduces the assumptions to the minimum. Device-independent tests are especially important within cryptographic scenarios and constitute, in a sense, the highest level of certification attainable with quantum theory. Here, we have reported the results of an experiment certifying for the first time a MIC-POVM for qubits following a device-independent protocol. For that, we have produced correlations between separated photons that, as we have proven, are only attributable to an informationally complete four-outcome measurement on qubits. Our results pave the way towards realistic applications requiring device-independent certification of MIC-POVMs, and show how very refined concepts of quantum information are now experimentally attainable and can transform communication and information processing technologies. 

\section*{Funding}

Swedish Research Council (Project No. 2017-04855); 
Knut and Alice Wallenberg Foundation, Sweden (Project ``Photonic Quantum Information''); 
MINECO-MICINN with FEDER funds (Project No.\ FIS2017-89609-P ``Quantum Tools for Information, Computation and Research''); 
Conserjer\'{\i}a de Conocimiento, Investigaci\'on y Universidad, Junta de Andaluc\'{\i}a and European Regional Development Fund (ERDF) (Ref.~SOMM17/6105/UGR);
National Science Centre (NCN) (Grant No.\ 2014/14/E/ST2/00020);
National Science Centre (NCN) (Grant No.\ 2018/30/Q/ST2/00625);
DS Programs of the Faculty of Electronics, Telecommunications and Informatics, Gda\'nsk University of Technology;
First TEAM (Grant No. First TEAM/2016-1/5).


\section*{Disclosures}

The authors declare no conflicts of interest.

\section*{}
See Supplemental Material for supporting content.


\bibliography{MIC_POVM-bibliography}

\newpage

\section{Supplemental material}
Here we provide supplementary information to the manuscript above. In particular, the following will be presented: details on the experimental realization of non-projective measurements, proof of information completeness, results of tomographic reconstruction of Alice's eight local states and details on error estimation.


\subsection{Experimental realization of Alice's POVM}\label{app:povm_real}
We will derive in this section the Kraus operators corresponding to the four outcomes of Alice's non-projective measurement (Eq.~(4) in main paper).
In order to get to them, we shall work with a four dimensional Hilbert space on Alice's side, which includes the usual polarization space (with basis vectors $\ket{H}$ and $\ket{V}$) and the additional path degree of freedom added by the Sagnac interferometer. Referring to Fig.~2 in the main paper, we denote by $\ket{a}$ the mode transmitted by the polarizing beam splitter (PBS) at the entrance of the interferometer, passing through lambda-half wave plate (HWP) H1, and transmitted again by the PBS. Counter-propagating to it, and going through HWP H2, is instead mode $\ket{b}$.

We can then describe any four-dimensional state $\ket{\Psi}_A$ as a vector with basis $\{ \ket{H}\ket{a},\ket{H}\ket{b},\ket{V}\ket{a},\ket{V}\ket{b}\}$, where each element refers to one polarization-path combined mode.
In this context, a PBS can be described as:
\begin{equation}
 U_{PBS} = \begin{pmatrix} 1 & 0 & 0 & 0\\ 0 & 0 & 0 & i\\ 0 & 0 & 1 & 0\\ 0 & i & 0 & 0\end{pmatrix} ,
\end{equation}
while HWP, lambda-quarter wave plates (QWP) and phase plates (PP) as:
\begin{equation}
\begin{aligned}
 U_{H}(\theta) & = \begin{pmatrix} \cos2\theta & \sin2\theta\\ \sin2\theta & -\cos2\theta \end{pmatrix}, \\ 
U_{Q}(\theta) & = \begin{pmatrix} \cos^2\theta+i\sin^2\theta & (1-i)\sin\theta \cos\theta\\ (1-i)\sin\theta \cos\theta & i\cos^2\theta+\sin^2\theta \end{pmatrix}, \\
U_{PP}(\theta) & = \begin{pmatrix} 1 & 0\\ 0 & e^{i\theta} \end{pmatrix} .
\end{aligned}
\end{equation}
The HWPs inside the interferometer act as:
\begin{equation}
\begin{aligned}
 U&_{H1H2} = U_{H}(\theta_{H1}) \oplus U_{H}(\theta_{H2}) = \\
 & \begin{pmatrix} \cos2\theta_{H1} & \sin2\theta_{H1} & 0 & 0\\ \sin2\theta_{H1} & -\cos2\theta_{H1} & 0 & 0\\ 0 & 0 & \cos2\theta_{H2} & \sin2\theta_{H2}\\ 0 & 0 & \sin2\theta_{H2} & -\cos2\theta_{H2}\end{pmatrix}
\end{aligned}
\end{equation}
and the whole interferometer is thus given by $U_{int}(\theta_{H1},\theta_{H2})=U_{PBS} U_{H1H2} U_{PBS}$.

After the Sagnac interferometer, each of the two output paths includes a combination of PP, HWP and QWP, PBS and two single photon detectors.
Referring to Fig.~2 in the main paper, we can then express the transformations in polarization space for outcomes $j=\{1,2,3,4\}$ in terms of Kraus operators as:
\begin{equation} \label{eq:povm_kraus}
\begin{aligned}
A_1 &=\bra{a} U_{PBS} U_{Q}(\theta_{Qa}) U_{H}(\theta_{Ha}) U_{PP}(\theta_{Pa}) \ket{a} \bra{a} U_{int} \ket{a},\\
A_2 &=\bra{a} U_{PBS} U_{Q}(\theta_{Qb}) U_{H}(\theta_{Hb}) U_{PP}(\theta_{Pb}) \ket{a} \bra{b} U_{int} \ket{a},\\
A_3 &=\bra{b} U_{PBS} U_{Q}(\theta_{Qb}) U_{H}(\theta_{Hb}) U_{PP}(\theta_{Pb}) \ket{a} \bra{b} U_{int} \ket{a},\\
A_4 &=\bra{b} U_{PBS} U_{Q}(\theta_{Qa}) U_{H}(\theta_{Ha}) U_{PP}(\theta_{Pa}) \ket{a} \bra{a} U_{int} \ket{a},\\
\end{aligned}
\end{equation}
so that Alice's qubit undergoes the operation $\ket{\psi}_A \rightarrow A_j \ket{\psi}_A$, and each of her non-projective measurement operators (in Eq.~(4) in the main paper) is described by $A_{4,j} = A_j^\dag A_j$.
As a side note, while a combination of HWP and QWP at each interferometer output would in principle be sufficient, adding the PPs allows for ``standard'' $\sigma_x$ and $\sigma_y$ measurement settings to be used, together with a fixed phase given by the PPs. Moreover, since the relative phase between the interferometer's arms is fixed but unknown, the additional PP simplifies the experimental task of compensating for this additional phase.

Because of the effectively redundant PPs, there are several combinations of settings that lead to optimal violation of Eq.~(5). In our experimental realization, we used the following: $\theta_{H1} = 31.32\degree$, $\theta_{H2} = 13.68\degree$, $\theta_{Pa} = 45\degree$, $\theta_{Ha} = 0\degree$, $\theta_{Qa} = 45\degree$, $\theta_{Pb} = 135\degree$, $\theta_{Hb} = 22.5\degree$, $\theta_{Qb} = 0\degree$.

\subsection{Information completeness of 4-outcome POVMs in dimension 2}

We will now show that if a $4$-outcome POVM is implemented in dimension $2$, then the information retrieved from the quantum system is complete.

Note that a system of dimension $2$ contains $3$ independent parameters. Thus, in order to show that a $4$-outcome POVM obtains these parameters we need to show that it consists of $3$ linearly independent operators.

Let us assume that a POVM contains only $2$ linearly independent operators. Without loss of generality we can write it in one of the forms:
\begin{equation}
	\label{eq:operPlus}
	\left( A, B, \alpha A + \beta B, \mathbbm{1} - (1+\alpha) A - (1+\beta) B \right),
\end{equation}
or
\begin{equation}
	\label{eq:operMinus}
	\left( A, B, \alpha A - \beta B, \mathbbm{1} - (1+\alpha) A - (1-\beta) B \right),
\end{equation}
with $\alpha,\beta \geq 0$. We now show that POVMs in~\eqref{eq:operPlus} and~\eqref{eq:operMinus} can be expressed as a convex combination of POVMs with at most $3$ outcomes.

Indeed, for~\eqref{eq:operPlus} we have:
\begin{equation}
	\begin{aligned}
		&\left( A, B, \alpha A + \beta B, \mathbbm{1} - (1+\alpha) A - (1+\beta) B \right) \\
		& \quad = \frac{1}{(1+\alpha)(1+\beta)} \left((1+\alpha)A, (1+\beta)B, 0, R \right) \\
		& \quad + \frac{\alpha}{(1+\alpha)(1+\beta)} \left(0, (1+\alpha)A, (1+\beta)B, R \right) \\
		& \quad + \frac{\beta}{(1+\alpha)(1+\beta)} \left((1+\alpha)A, 0, (1+\beta)B, R \right) \\
		& \quad + \frac{\alpha \beta}{(1+\alpha)(1+\beta)} \left(0, 0, (1+\alpha)A + (1+\beta)B, R \right),
	\end{aligned}
\end{equation}
where $R = \mathbbm{1} - (1+\alpha)A - (1+\beta)B$.

Similarly, for~\eqref{eq:operMinus} we have:
\begin{equation}
	\begin{aligned}
		&\left( A, B, \alpha A - \beta B, \mathbbm{1} - (1+\alpha) A - (1-\beta) B \right) \\
		& \quad = \frac{\alpha^2}{(\alpha+\beta)(1+\alpha)} \left( 0, \left(1+\frac{\beta}{\alpha}\right) B, \right. \\
		& \left. \hspace{3.1cm} \left(1+\frac{1}{\alpha}\right) (\alpha A - \beta B), R' \right) \\
		& \quad + \frac{\alpha}{(\alpha+\beta)(1+\alpha)} \left( \left(1+\frac{1}{\alpha}\right) (\alpha A - \beta B), \right. \\
		& \left. \hspace{3.1cm} \left(1+\frac{\beta}{\alpha}\right) B, 0, R' \right) \\
		& \quad + \frac{\alpha \beta}{(\alpha+\beta)(1+\alpha)} \left( \left(1+\frac{\beta}{\alpha}\right) B, 0, \right. \\
		& \left. \hspace{3.1cm} \left(1+\frac{1}{\alpha}\right) (\alpha A - \beta B), R' \right) \\
		& \quad + \frac{\beta}{(\alpha+\beta)(1+\alpha)} \left( (1+\alpha) A + (1-\beta) B, 0, 0, R' \right),
	\end{aligned}
\end{equation}
where $R' = \mathbbm{1} - (1+\alpha) A - (1-\beta) B$.

\subsection{Detection efficiency}

In order to calculate critical detection efficiency of the modified elegant Bell expression (Eq.~(7) in the main paper), we first calculated its local hidden variables bound. To this end we enumerated all possible deterministic strategies. This revealed that the strategy assigning outcomes $+$ for Alice's measurements $1$ and $3$, $-$ for Alice's measurement $2$, $2$ for the POVM measurement, $+$ for Bob's measurements $1$, $2$ and $4$ and $-$ for his measurement $3$ yields the value $6.1652$.

For detection efficiency calculations we assumed that, in post-processing of the experimental data whenever no-click events occur, an outcome from the optimal LHV strategy above is assigned. The critical detection efficiency calculated using the method in Ref.~\cite{CP18} is equal to $0.9439$.

\subsection{Tomographic reconstruction of Alice's eight local states}\label{app:tomo_vectors}
In Table~\ref{tab:vector_results} we report the eight qubit states, in Bloch vector notation, reconstructed using both standard projective tomography ($\sigma_x,\sigma_y,\sigma_z$), and with single-setting MIC-POVM tomography. These states are the local states of Alice's qubit, conditioned on Bob's measurement settings and outcomes. The pairwise fidelity is also reported.


\begin{table*}[htb]
	\begin{center}
		\begin{tabular}{ccc}
			\hline
			\qquad Projective tomography \qquad & \qquad SIC-POVM tomography \qquad & \qquad Fidelity \qquad \\ \hline
			$\begin{pmatrix}0.561 & 0.601 & 0.570\end{pmatrix}$ & $\begin{pmatrix}0.544 & 0.508 & 0.668\end{pmatrix}$ & 0.995$^{+0.004}_{-0.004}$\\
			$\begin{pmatrix}0.570 & -0.589 & -0.574\end{pmatrix}$ & $\begin{pmatrix}0.506 & -0.681 & -0.530\end{pmatrix}$ & 0.996$^{+0.003}_{-0.004}$\\
			$\begin{pmatrix}-0.572 & 0.525 & -0.630\end{pmatrix}$ & $\begin{pmatrix}-0.572 & 0.525 & -0.630\end{pmatrix}$ & 0.997$^{+0.002}_{-0.003}$\\
			$\begin{pmatrix}-0.551 & -0.590 & 0.591\end{pmatrix}$ & $\begin{pmatrix}-0.526 & -0.695 & 0.490\end{pmatrix}$ & 0.995$^{+0.002}_{-0.005}$\\
			$\begin{pmatrix}-0.548 & -0.581 & -0.601\end{pmatrix}$ & $\begin{pmatrix}-0.518 & -0.574 & -0.634\end{pmatrix}$ & 0.999$^{+0.001}_{-0.001}$\\
			$\begin{pmatrix}-0.577 & 0.611 & 0.541\end{pmatrix}$ & $\begin{pmatrix}-0.658 & 0.586 & 0.474\end{pmatrix}$ & 0.997$^{+0.002}_{-0.002}$\\
			$\begin{pmatrix}0.588 & -0.532 & 0.610\end{pmatrix}$ & $\begin{pmatrix}0.560 & -0.603 & 0.569\end{pmatrix}$ & 0.998$^{+0.001}_{-0.001}$\\
			$\begin{pmatrix}0.540 & 0.573 & -0.616\end{pmatrix}$ & $\begin{pmatrix}0.480 & 0.570 & -0.667\end{pmatrix}$ & 0.998$^{+0.001}_{-0.002}$\\ \hline
		\end{tabular}
	\end{center}
\caption{\label{tab:vector_results}Tomographic reconstruction of the states depicted in Fig.~4 in the main text, using both projective and MIC-POVM tomography, and their pairwise fidelities. Uncertainties represent $15.9\%$ and $84.1\%$ quantiles of the respective results' distributions.}
\end{table*}


\subsection{Error estimation}\label{app:errors}
Here we provide a more comprehensive description of the errors considered in the experiment.

\subsubsection{Counting statistics}
Whenever (coincident) events with a constant rate are counted for some amount of time, the distribution of the final amount is in very good approximation Poissonian. We therefore considered all our empirical counts to have an uncertainty equal to their square root, and propagated it in the results. This is, by far, the predominant contribution to the final uncertainties in our experiment, giving errors of the order of $2\cdot10^{-3}$ and $10^{-4}$ on each $E_{ab}$ and $P(a=i,b=+1|x=4,y=i)$ term, respectively.


\subsubsection{Motor precision}
All measurement wave plates were rotated by motorized mounts controlled by a computer. The step motors have a precision equivalent to $0.02 \degree$. This results in errors of the same order of the Poissonian ones. In order to reduce their contribution, each setting was repeated 23 times, therefore decreasing the uncertainties by almost a factor of 5.


\subsubsection{Detector dark counts}
Each of the single photon detectors used in the measurements have dark count rates of about 500 detections per second. The chances of a coincident event stemming from a true detection and a dark count, with the rates used, was as low as $10^{-11}$, thus negligible.


\subsubsection{Higher order down-conversion events}
The rate of accidental coincidences $ac_{m,ij}$ coming from multiple down-conversion events in a single pulse, for measurement setting $m$ and detectors $(i,j)$, can be estimated with the formula
\begin{equation}
 ac_{m,ij} = \frac{S_{m,i} S_{m,j} \Delta t}{T},
\end{equation}
where $S_{m,k}$ are the total (single) events on detector $k$ during measurement time $T$, when coincidence windows of length $\Delta t$ are used. While the resulting rates are fairly minimal (of the order of $10^{-3}$ events per second), they can still worsen, although slightly, the results obtained. Since the DI certification protocol can, in principle, work even if the state and measurements are not characterized, we chose not to correct our evaluations for this type of error. 
In the case of the full state tomography, the derived fidelity of $99.6 \% $ (and corresponding uncertainty), did not change whether we took accidental counts into consideration or not, due to their very low rate.

\end{document}